\documentclass[useAMS,usenatbib,usegraphicx,letterpaper]{mn2e}

\usepackage{multirow} 
\usepackage{epsfig}
\usepackage{subfigure}

\newcommand{\etal}{et al.}

\def\mnras{MNRAS}
\def\apj{ApJ}
\def\apjs{ApJS}
\def\aap{A\&A}
\def\aj{AJ}

\title[An intermediate BH spin in SWIFT~J2127.4] {An
  intermediate black hole spin in the NLS1 galaxy SWIFT~J2127.4+5654:
  chaotic accretion or spin energy extraction?}

\author[G.\ Miniutti \etal]
       {G.~Miniutti$^{1}$\thanks{gminiutti@laeff.inta.es},
         F. Panessa$^2$, A. De Rosa$^2$, A.C. Fabian$^3$,
         A. Malizia$^4$, M. Molina$^5$, \newauthor J.M. Miller$^6$ and
         S. Vaughan$^7$ \\ \\ $^1$ LAEX, Centro de Astrobiologia
         (CSIC--INTA); LAEFF, P.O. Box 78, E--28691, Villanueva de la
         Ca\~{n}ada, Madrid, Spain \\ $^2$ IASF/INAF, via del Fosso del
         Cavaliere 100, I--00133 Roma, Italy\\ $^3$ Institute of Astronomy,
         Madingley Road, CB3 0HA Cambridge \\ $^4$ IASF/INAF, via
         Gobetti 101, I--40129 Bologna, Italy \\ $^5$ IASF/INAF, via Bassini
         15, I--20133 Milano, Italy \\ $^6$ Department of Astronomy,
         University of Michigan, 500 Church Street, Ann Arbor, MI
         48109, USA \\ $^7$ X--ray Astronomy Group, University of
         Leicester, LE1 7RH Leicester\\ }

\pagerange{\pageref{firstpage}--\pageref{lastpage}}
\pubyear{2001}

\usepackage{times}

\begin{document}

\label{firstpage}

\maketitle

\begin{abstract}
We have observed the hard X--ray selected Narrow--Line Seyfert~1
galaxy SWIFT~J2127.4+5654 with {\it Suzaku}. We report the detection
of a broad relativistic iron emission line from the inner accretion
disc. Partial covering by either neutral or partially ionized gas
cannot reproduce the spectral shape and, even if its presence is
assumed, it does not significantly change the inferred broad line
parameters. By assuming that the inner edge of the accretion disc
corresponds to the innermost stable circular orbit of the black hole
spacetime, the line profile enables us to measure a black hole spin
$a= 0.6\pm 0.2$. However, a non--rotating Schwarzschild spacetime is
excluded at just above the 3$\sigma$ level, while a maximally spinning
Kerr black hole is excluded at the $\sim$5$\sigma$ level. The
intermediate spin we measure may indicate that accretion--driven black
hole growth in this source proceeds through short--lived episodes with
chaotic angular momentum alignment between the disc and the hole
rather than via prolonged accretion. The measured steep emissivity
index ($q\simeq 5$) constrains the irradiating X--ray source to be
very centrally concentrated. Light bending may help focus the primary
emission towards the innermost accretion disc, thus steepening the
irradiation profile. On the other hand, steep profiles can also be
reached if magnetic extraction of the hole rotational energy is at
work. If this is the case, the interplay between accretion (spinning
up the black hole) and rotational energy extraction (spinning it down)
forces the hole to reach an equilibrium spin value which, under
standard assumptions, is remarkably consistent with our
measurement. Rotational energy extraction would then be able to
simultaneously account for the intermediate spin and steep emissivity
profile we infer from our spectral analysis without the need to invoke
chaotic accretion episodes. We also report tentative evidence for
short timescale line profile variability. The relatively low
statistical significance of the variability (about 98 per cent
confidence level) prevents us from drawing any firm conclusions which
must be deferred to future observations.
\end{abstract}


\section{Introduction}

X--ray irradiation of dense and relatively cold material in the
vicinity of accreting black holes can produce an X--ray reflection
spectrum characterized by fluorescent emission lines (most notably in
the iron K regime around 6--7~keV), a Compton hump around
$\sim$20--30~keV, and a soft excess below 1--2~keV. If the reflection
spectrum is emitted from the inner accretion disc, the whole spectrum
is distorted by relativistic effects. Relativistic emission line
profiles bear the imprint of the strong Doppler and gravitational
energy shifts (plus aberration and light bending) which are intrinsic
to the inner disc, where velocities and gravity are the largest. Given
that the iron (Fe) K emission line is the strongest in the X--ray
regime, relativistic Fe line studies can serve to probe the
relativistic spacetime close to black holes (Fabian 1989; Fabian et
al. 2000; Reynolds \& Nowak 2003; Fabian \& Miniutti 2009).

The astrophysical black hole spacetime only depends on black hole mass
and spin. While the black hole mass sets the scale of the system, the
spin qualitatively changes the nature of the spacetime and can thus be
regarded as the most crucial parameter. X--ray observations of
accreting Galactic black holes and of Active Galactic Nuclei (AGN) are
now starting to attack the difficult process of measuring  black hole spin
through relativistic Fe line diagnostics (e.g. Miller et al. 2009;
Brenneman \& Reynolds 2006). Measuring such an important parameter in
large and statistically meaningful samples of accreting sources in the
future is crucial to provide answers to several open questions in the
field. For instance, such studies will allow observers to test whether a non
zero black hole spin is a necessary condition to launch relativistic
jets (see e.g. Sikora, Stawarz \& Lasota 2007). In Galactic sources,
since accretion from the companion star is unlikely to significantly
modify the spin imprinted at birth (Volonteri et al. 2005), a measure
of the black hole rotation will provide crucial clues on the dynamics
and black hole formation during the supernova explosion. On the other
hand, the spins of supermassive black holes in AGN bear the imprints
of the hole growth and evolution, and of the relative importance of i)
mergers and accretion and ii) prolonged versus chaotic/episodic
accretion in setting up the final black hole mass (e.g. Volonteri et
al. 2005; Berti \& Volonteri 2008; King, Pringle \& Hofmann 2008).

Here we report results from a {\it{Suzaku}} observation of the bright
Narrow--Line Seyfert~1 galaxy SWIFT~J2127.4+5654, previously detected
in  hard X--rays by {\it Swift}/BAT and {\it INTEGRAL}/IBIS and in
the soft X--ray band by the {\it{Swift}}/XRT.

\section{SWIFT~J2127.4+5654}

SWIFT~J2127.4+5654 (a.k.a. IGR~J21277+5656) is a low Galactic latitude
($\sim$4.4 degrees) hard X--ray source detected with {\it{Swift}}/BAT
(Tueller et al. 2005) and later identified as a Narrow--Line Seyfert 1
(NLS1) galaxy at redshift 0.0147 based on the observed H$\alpha$ FWHM
of $\sim$~1180~km/s (Halpern 2006). Subsequent work by Malizia et
al. (2008) supported the NLS1 classification by obtaining, despite a
slightly larger H$\alpha$ FWHM of $\sim$~2000~km/s, a relatively low
[{O\,\textsc{iii}}]/H$_\beta$ ratio of $0.72\pm 0.05$ and
significantly enhanced {Fe\,\textsc{ii}} emission
({Fe\,\textsc{ii}}/H$_\beta =1.3\pm 0.2$). Some radio-loud AGN exhibit
optical spectra which display narrow permitted lines and weak
[{O\,\textsc{iii}}] emission and are so potentially misclassified as
NLS1 galaxies. SWIFT~J2127.4+5654 has a 20~cm flux of 6.4~mJy (Condon
et al. 1998). The source appears compact, with no visible elongation
in the 1.4~GHz NRAO VLA Sky Survey (NVSS), and is certainly not bright
enough to be classified as a radio--loud source, so that the NLS1
classification of SWIFT~J2127.4+5654 appears robust (we note anyway
that the NLS1 nature of AGN and radio--loudness are not necessarily
mutually exclusive, see Komossa et al. 2006).

The source was detected in the hard band with {\it INTEGRAL}/IBIS
and, although no spectral information can be extracted from a single
IBIS pointing, an averaged spectrum was obtained by Malizia et
al. (2008) in the 17--100~keV band by summing all the available
on--source exposures, resulting in a steep spectrum ($\Gamma$ in the
range 2.4--3) and a hard X--ray flux of $2.46 \times
10^{-11}$~erg~s$^{-1}$~cm$^{-2}$ in the 20--100~keV band. The
{\it{Swift}}/XRT data (two observations for a total of $\sim$10~ks)
have been analysed by Malizia et al. (2008) in conjunction with the
{\it INTEGRAL}/IBIS averaged spectrum. The main result of this analysis
is that the softer X--rays probed by {\it{Swift}}/XRT exhibit a much
flatter and more typical spectral slope of $\Gamma \sim 1.9$
suggesting a more complex broadband spectral shape than a simple power
law. Although the averaged {\it INTEGRAL}/IBIS high--energy data may
be affected by variability, the analysis suggests the presence of a
spectral break or of an exponential cut--off in the range of
20--50~keV. If the continuum is assumed to originate from inverse
Compton of the soft disc photons in a hot corona, a low energy
cut--off in that energy range implies a low temperature of the
Comptonizing plasma (about 10--20~keV). On the other hand, other
interpretations are viable: for instance the broadband X--ray spectrum
may be characterized by the presence of a relatively steep continuum
plus a harder reflection component and/or absorption leading to the
observed apparent spectral break in the hard X--ray band.

\section{The {\it Suzaku} observations}

{\it Suzaku} observed SWIFT~J2127.4+5654 on 2007 December 9 for a
total net exposure of 92~ks. SWIFT~J2127.4+5654 was placed at the HXD
nominal position. Here we consider data from the XIS (Koyama et
al. 2007) and HXD/PIN (Takahashi et al. 2007) detectors. During our
observation, the front--illuminated CCD XIS~0 was not operating
properly and, after the loss of XIS~2, the only available CCD spectra
are those provided by the back--illuminated XIS~1 and the
front--illuminated XIS~3.  We analyse the v2.2 data and we use the
HEASOFT 6.5 distribution and associated CALDB released on
2008--08-11. All spectra we use are grouped to a minimum of 50
counts per bin. We use the $\chi^2$ minimization technique and we
quote 90 per cent errors on model parameters unless specified
otherwise. Spectral analysis is performed with the XSPEC v~12.5
software (Arnaud 1996; Dorman \& Arnaud 2001). Luminosities are
computed by assuming a $\Lambda$CDM cosmology with $H_0 =
70$~km~s$^{-1}$, $\Omega_\Lambda = 0.73$ and $\Omega_{\rm M}=0.27$.

\section{The {\it{Suzaku}} view of SWIFT~J2127.4+5654}

\begin{figure}
\begin{center}
\includegraphics[width=0.31\textwidth,height=0.45\textwidth,angle=-90]{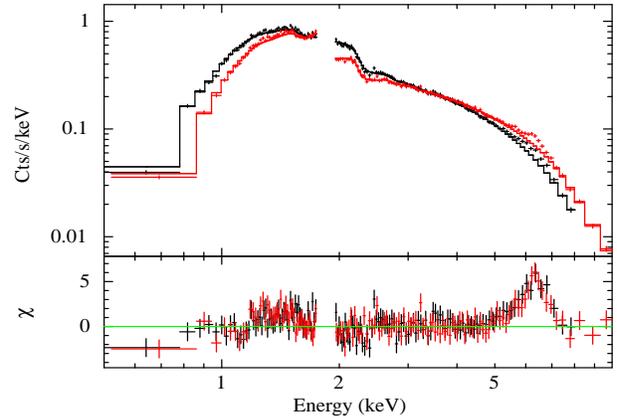}
\caption{XIS~1 and XIS~3 spectra, model, and residuals for a
  simple absorbed power law fit. The data are re--binned for visual
  clarity only.}
\label{ratio1}
\end{center}
\end{figure}

We first consider the XIS~1 and XIS~3 time--averaged spectra in the
0.5--10~keV (XIS~3) and 0.5--8~keV (XIS~1) band, ignoring the
1.75--1.95~keV energy range because of remaining uncertainties in the
instrumental Si edge region and because we see in the data a suspect
sharp spectral feature not accounted for by the current
calibration. The spectra are fitted jointly with a simple absorbed
power law model. We use a $z=0$ photoelectric absorption model
({\small{PHABS}}) with column density fixed at the galactic value
($\sim 7.7 \times 10^{21}$cm$^{-2}$, Kalberla et al. 2005) and a
further intrinsic column N$_{\rm H}^{\rm{z}}$ at the redshift of the
source (using the {\small{ZPHABS}} model). Abundances are set to
Wilms, Allen \& McCray (2000) and we use photoelectric absorption
cross--sections from Balucinska-Church \& McCammon (1992). The model
provides a reasonable description of the data ($2881$ for 2670 degrees
of freedom, dof) and we measure N$_{\rm H}^{\rm z} = 2.9 \pm 0.1\times
10^{21}$cm$^{-2}$ with photon index $\Gamma =1.98\pm 0.02$. Replacing
the intrinsic neutral absorber with an {\small{XSTAR}}--based ionized
one\footnote{We use the {\small{XSTAR}} code included in the HEASOFT
  6.5 distribution to build an ionized absorber {\small{XSPEC}} table
  model (Bautista \& Kallman 2001). We assume a $\Gamma=2$ continuum
  between 0.1 and 1000~Ry, and a turbulent velocity of
  200~km~s$^{-1}$. The column density ranges between $0.5$ and
  $50\times 10^{21}$~cm$^{-2}$, while $\log\xi$ ranges between $-1$
  and $2.5$.} does not improve the fit and the ionization parameter is
constrained to be $\log\xi \leq -0.5$.

Our spectral best--fitting model is a fair statistical description of the data,
but leaves however clear residuals as shown in Fig.~\ref{ratio1}. The
most important residuals are in the Fe K band around 6~keV and
suggest the presence of a broad Fe line. On the other hand, some
spectral curvature is also suggested by the soft X--ray data below
about 2~keV possibly indicating a small soft excess (where ``small''
should however take into account the high observed column density
which would mask anyway a larger soft excess). Below 1~keV we also
note the possible presence of excess absorption which is not accounted
for by the current spectral model.

We first consider the possible presence of a soft excess by
introducing a multi--color blackbody model ({\small{DISKBB}}, see
Makishima et al. 1986). The quality of the fit improves significantly
($\chi^2=2755$ for 2668 dof, i.e. $\Delta\chi^2 = -126$ for
$\Delta$dof$=-2$) and we obtain an inner disc temperature of $340\pm
40$~eV with photon index $\Gamma = 1.92 \pm 0.03$ and N$_{\rm
  H}^{\rm{z}}= 3.8 \pm 0.2\times 10^{21}$~cm$^{-2}$. We point out that
such a high blackbody temperature is unphysical and cannot be
associated with thermal emission from the accretion disc (which peaks
in the UV) so that the blackbody model has to be considered as
phenomenological. Given the different soft spectral shape, we
re--investigate the presence of an ionized intrinsic absorber. Using
the same {\small{XSTAR}} model as discussed above, we replace the
intrinsic neutral absorber with an ionized one, and we obtain a
statistical improvement ($\chi^2=2732$ for 2667 dof) with low
ionization parameter $\log\xi = 0.15 \pm 0.10$ and N$_{\rm
  H}^{\rm{z}}= 6.6 \pm 0.5\times 10^{21}$~cm$^{-2}$. The disc
blackbody component has now a temperature of $\sim 240$~eV (still too
hot to represent the high--energy tail of the disc thermal emission)
and it contributes by about 20 per cent to the total unabsorbed
luminosity in the 0.5--2~keV band, in line with the typical $\sim$30
per cent soft excess contribution in Seyfert 1 galaxies and type I
quasars, e.g. Miniutti et al. (2009).

\begin{table}
\begin{center}
  \caption{Best--fitting parameters for the joint fit of XIS~1 and
    XIS~3 data. $N_H^{\rm{z}}$ and $\log\xi^{\rm{z}}$ refer to the
    intrinsic ionized absorber at the redshift of the source, while
    the subscripts BL and NL refer to the broad and narrow Gaussian
    emission lines in the Fe K band.}
\begin{tabular}{lc}          
  \hline
  $\Gamma$ & $2.06 \pm 0.03 $ {\vspace{0.05cm}}\\
  $N_H^{\rm{z},a}$ [cm$^{-2}$]&  $6.6 \pm 0.5$ {\vspace{0.05cm}}\\
  $\log\xi^{\rm{z},a}$ &  $0.15 \pm 0.10$ {\vspace{0.05cm}}\\
  $kT_{\rm inn}$ [eV]&  $240\pm 20$ {\vspace{0.05cm}}\\
  $E_{\rm{BL}}$ [keV]&  $6.35\pm 0.10$ {\vspace{0.05cm}}\\
  $\sigma_{\rm{BL}}$ [keV]& $0.53\pm 0.10$ {\vspace{0.05cm}}\\
  $EW_{\rm{BL}}$ [eV]&  $220\pm 50$ {\vspace{0.05cm}}\\
  $EW_{\rm{NL}}$ [eV]&  $15\pm 10$ {\vspace{0.05cm}}\\
  $F_X^b$ [erg~s$^{-1}$~cm$^{-2}$]& $3.29\pm 0.03$ {\vspace{0.05cm}}\\ 
  $L_X^c$ [erg~s$^{-1}$]& $1.74\pm 0.01$ {\vspace{0.05cm}}\\ 
  $L^d_{\rm Bol}/L_{\rm Edd}$ & $\simeq 0.18$ {\vspace{0.05cm}}\\ 
  $\chi^2 /dof$ & 2563/2663 \\
\hline
\end{tabular}
\end{center}
\label{tab1}
{\footnotesize{$^a$ $N_H^{\rm{z}}$ denotes the column density of the
    intrinsic absorber at $z=0.0147$ in units of $10^{21}$~cm$^{-2}$,
    while its ionization parameter ($\xi^{\rm z}$) is in units of
    erg~cm~s$^{-1}$; $^b$ absorbed X--ray flux in the 2--10~keV band
    in units of $10^{-11}$~erg~s$^{-1}$~cm$^{-2}$; $^c$ unabsorbed
    X--ray luminosity in the 2--10~keV band in units of
    $10^{43}$~erg~s$^{-1}$; $^{d}$ We compute $L_{\rm Bol}$ from the
    2--10~keV X--ray luminosity applying the Marconi et al. (2004)
    bolometric correction. We assume a black hole mass of $1.5\times
    10^7~M_\odot$ (Malizia et al. (2008).}}
\end{table}

The above modelling provides a good description of the soft X--ray
data, but the main positive residuals in Fig.~\ref{ratio1} in the Fe K
band are not accounted for yet. To model the residuals around 6~keV we
introduce a Gaussian emission line with energy, width, and
normalization free to vary. The Gaussian emission line provides a
further very significant improvement of the statistical quality of the
fit, with a $\Delta\chi^2 = -165$ for $\Delta$dof$= -3$ ($\chi^2=2567$
for 2664 dof). The Gaussian line has an energy of $6.35\pm 0.10$~keV,
consistent with neutral Fe K$\alpha$ emission, and is resolved with a
relatively large width of $\sigma=0.53\pm 0.10$~keV.

Given that a narrow Fe K$\alpha$ line (most likely originating from
distant reflection) is almost ubiquitous in the X--ray spectra of AGN
(e.g. Bianchi et al. 2009), we also add a narrow Gaussian emission
line with energy $E\equiv 6.4$~keV and width $\sigma\equiv 1$~eV. The
narrow Fe line does not improve very significantly the quality of the
fit ($\Delta\chi^2 = -4$ for 1 dof less than before) but we keep it
for consistency, measuring a line intensity of
$4.6^{+4.0}_{-4.1}\times 10^{-6}$~ph~s$^{-1}$~cm$^{-2}$.  The broad
line has an equivalent width (EW) of $220\pm 50$~eV, while the narrow
Fe line EW is $15\pm 10$~eV. Adding narrow Fe~\textsc{xv} and/or
Fe~\textsc{xvi} emission/absorption lines in the Fe K band does not
improve the fit significantly and does not change the broad Fe
K$\alpha$ line parameters at all.
The best--fitting parameters for our final model are given in Table~1.

Broad Fe lines are sometimes claimed to be a spurious effect due to
complex absorption (e.g. Turner \& Miller 2009). In particular,
partial covering scenarios potentially introduce hard X--ray spectral
curvature which can be misinterpreted as a broad Fe line. As a sanity
check, we then add a partial covering ionized absorber to our model
(the {\small{ZXIPCF}} model in {\small{XSPEC}}). However, the partial
covering absorber does not improve the statistical quality of the fit
($\chi^2=2561$ for 2660 dof) and no constraints other than a small
covering fraction of less than 20 per cent can be obtained. We obtain
the same unconstrained result if a neutral partial coverer is used
instead. Moreover, no significant variation in the broad line
parameters is seen in either case (new parameters are E$=6.36\pm
0.11$~keV, $\sigma = 0.51\pm 0.12$~eV for a line EW$\simeq 210$~eV for
both attempts). Replacing the broad Fe line with the ionized or
neutral partial covering model and re--fitting the data produces a
much worse description of the data (the minimum $\Delta\chi^2$ being
$\Delta\chi^2 = +56$ with the neutral partial coverer with respect to
the best--fit with broad line and no partial covering). We then
conclude that the detection of a broad Fe line in the X--ray spectrum
of SWIFT~J2127.4+5654 is robust against partial covering scenarios
irrespectively of the absorber ionization state.

\subsection{The relativistic Fe line: a probe of black hole spin}

Having established that a broad Fe line is present in the data with
high statistical significance, we now investigate its nature in some
more detail. The most likely origin for the broad Fe line is
reflection off the inner accretion disc. Relativistic line models
depend on several parameters, namely the inner and outer disc radii,
the disc--observer inclination, and the shape of the emissivity
profile (plus line energy and normalization). The emissivity is
generally assumed to be axisymmetric and with a power law ($\epsilon
\propto r^{-q}$, where $q$ is the emissivity index) or broken power
law shape.

The inner disc radius is of particular interest. Under the hypothesis
that the reflecting disc is truncated at the innermost stable circular
 orbit (ISCO) around the black hole (see Reynolds \& Fabian 2008
for a recent discussion), the inner disc radius only depends on the
black hole spin. Relativistic line models in which the black hole spin
is a variable free parameter are now available (the {\small{KYRLINE}}
and {\small{KERRDISK}} models; Dov{\v c}iak, Karas \& Yaqoob 2004 and
Brenneman \& Reynolds 2006 respectively). These models work under the
assumption that the inner disc radius is identified with the ISCO and
compute the orbital motion self--consistency (as opposed to the cases
of the {\small{DISKLINE}} and {\small{LAOR}} models which assume a
particular black hole spin, see Fabian et al. 1989 and Laor
1991). After checking that results from the {\small{KYRLINE}} and
{\small{KERRDISK}} models are consistent with each other in terms of
statistics, best--fitting values, and errors on the parameters, we
report results for the {\small{KYRLINE}} model only.

We replace the broad Gaussian Fe line of our previous model with the
{\small{KYRLINE}} relativistic line. We assume a neutral Fe line at
6.4~keV, as suggested by previous results (see Table~1), and we fix
the outer disc radius to 400~$r_g$, while black hole spin, emissivity
index $q$, and disc inclination $i$ are free to vary. We keep the
ionized absorber, disc blackbody, power law, and narrow Fe K line
components as before (all parameters free to vary except for the
narrow Fe line energy and width fixed at 6.4~keV and 1~eV
respectively). The Fe line model is better physically motivated than
the broad Gaussian used above, and provides a similar (although not a
better) statistical quality with $\chi^2/dof = 2561/2662$, to be
compared with $\chi^2/dof = 2563/2663$ when the broad Gaussian model
was used. We measure $q=5.2^{+1.8}_{-1.2}$, $i=46^\circ\pm 4^\circ$,
and a black hole spin of $a=0.66^{+0.12}_{-0.09}$.

\begin{figure}
\begin{center}
\includegraphics[width=0.31\textwidth,height=0.45\textwidth,angle=-90]{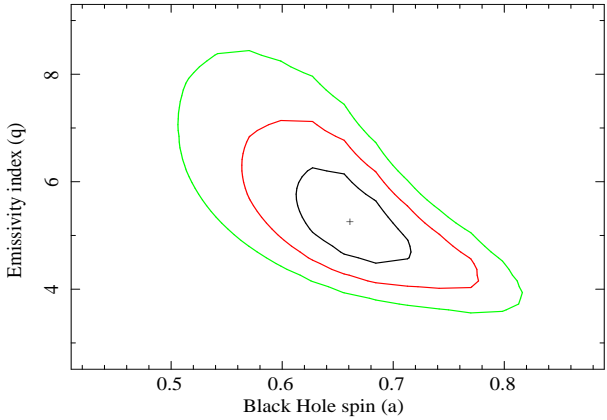}
\caption{$\Delta \chi^2 = 2.30$, $6.17$ and $11.83$ contours (corresponding to 
approximate 68.3, 95.4 and 99.7 per cent confidence regions) for the two
  most relevant relativistic blurring parameters, namely black hole
  spin ($a$) and emissivity index ($q$).  }
\label{spincont}
\end{center}
\end{figure}

In Fig.~\ref{spincont} we show $\Delta \chi^2$ contours for the
emissivity index $q$ and the black hole spin, demonstrating that the
parameter space is constrained. If we consider $\sim$3$\sigma$ errors
for the two parameters (i.e. $\Delta\chi^2=11.83$), the black hole
spin is constrained to be $a=0.66\pm 0.16$. Finally, we have
re--fitted the data by forcing the spin to be either $a=0$ or
$a=0.998$, corresponding to a Schwarzschild and a maximally spinning
Kerr spacetime respectively. We obtain $\Delta\chi^2 = +13$ for $a=0$
and $\Delta\chi^2 = +15$ for $a=0.998$, meaning that both extremes are
rejected at the $\sim$3$\sigma$ level only.

The emissivity index we measure is steep and suggests that Fe emission
is concentrated in the very inner regions of the disc. If the
irradiating X--ray source is point--like, ray--tracing studies in the
Kerr geometry reveal that the irradiating profile cannot be that steep
out to the largest radii (e.g. Miniutti \& Fabian 2004). A more
physically plausible irradiating profile can be approximated with a
broken power law with a steep index out to a break radius, and
reaching its asymptotic al $r^{-3}$ behaviour at larger radii. Such
emissivity profile was indeed used by several authors to model the
highest signal--to--noise relativistic Fe line known so far in AGN,
that of MCG--6-30-15 (Fabian et al. 2002; , Brenneman \& Reynolds
2006; Miniutti et al. 2007). We then try to fit the data by
using the {\small{KERRDISK}} model which allows the user to define a
broken power law emissivity. We could not find any statistical
improvement by using such model, and the outer emissivity index is
always consistent with the inner one. A solution with $q_{\rm{in}}\sim
5$ out to a break radius of $5-10~r_g$ and $q_{\rm{out}}\equiv 3$ at
larger radii is however statistically equivalent to the one presented
above and possibly more physically motivated. Since this solution
cannot be explored in detail due to signal--to--noise limitations, we
prefer to continue our analysis with the simpler non--broken power law
emissivity profile discussed above.

The relativistic line model best--fitting parameters we infer
should be taken with care. In fact, the broad Fe line must be
associated with a whole X--ray reflection continuum which, by changing
the underlying spectral shape, may affect the derived parameters
(certainly reducing the line EW). Since the reflection continuum is
characterized by a Compton hump showing up in the hard X--rays around
20~keV, we proceed to include the HXD data in the attempt to build a
more self--consistent model.

\begin{figure}
\begin{center}
\includegraphics[width=0.31\textwidth,height=0.45\textwidth,angle=-90]{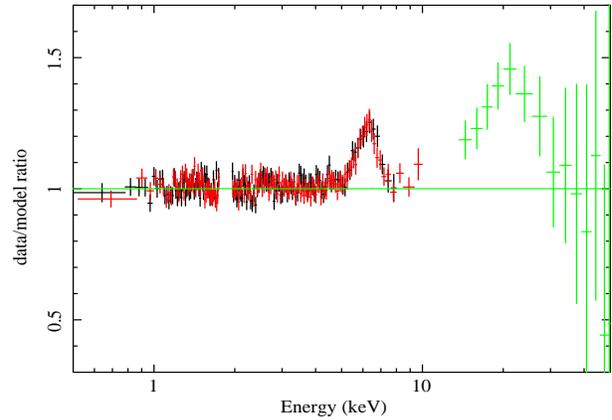}
\caption{The best--fitting continuum model is applied to the broadband
data. The Fe K band region has been ignored in the fit. Beside the
broad Fe line, the ratio highlights also positive residuals around
20~keV which indicate the likely presence of a Compton hump associated
with the reflection continuum.}
\label{extrap}
\end{center}
\end{figure}

\subsection{Including the high--energy {\it HXD/PIN} data}

{\it Suzaku}'s HXD and, in particular, the PIN detector is well suited
to investigate the presence of a reflection component in the spectrum
because of the good sensitivity up to 40--50~keV.  We thus consider
the 14--50~keV PIN data, and we introduce a cross--normalization
factor of $1.17\pm 0.02$ between the PIN and the XIS~3, rescaling the
recommended value\footnote{See Ishida et al., 2008--06--26,
  JX--ISAS--SUZAKU--MEMO--2008-06.} of $1.181\pm 0.016$ which is valid
for the XIS~0 (not available here) at the HXD nominal position taking
into account the XIS~3 to XIS~0 cross--normalization constant of
$1.009\pm 0.011$.  In Fig.~\ref{extrap} we show the data to model
ratio when the previous continuum model is applied to the broadband
data. The residuals clearly show, besides the broad Fe line, positive
residuals in the PIN band, most likely associated with the expected
Compton hump.

To test the Compton hump nature of the hard X--ray residuals, we
replace the power law continuum of our previous model with a power law
pus reflection continuum model, assuming reflection off a neutral slab
of gas (the {\small{PEXRAV}} model). We force the metal abundances to
be solar, and we fix the inclination of the reflector to $46^\circ$,
as suggested by the relativistic line fits discussed above. Although
the reflection model is not self--consistent because it does not
include line emission, it is useful to constrain the strength of the
reflection continuum and to search for a high energy cut--off below
$\sim$~50~keV as suggested by Malizia et al. (2008) in their analysis
of the combined {\it{Swift}} and {\it{INTEGRAL}} data.

We obtain a very good description of the broadband data ($\chi^2=2680$
for 2755 dof) and the hard residuals shown in Fig.~\ref{extrap}
disappear. The photon index is in line with our previous result ($\Gamma =
2.10 \pm 0.06$) and the relative contribution of the reflection
component (i.e. the reflection fraction $R$) is measured to be
$R=1.0^{+0.5}_{-0.4}$. As for the high energy cut-off at energy
$E_{\rm c}$, we obtain $E_c \geq 35$~keV, only marginally consistent
with the result by Malizia et al (2008). As expected, the inclusion of
the reflection continuum reduces the broad Fe line EW (now $\sim
150$~eV), while the narrow Fe line EW is $\sim 16$~eV making the total
EW associated with the Fe lines broadly consistent with the
theoretical expectations for solar Fe abundance in the case of neutral
matter (e.g. George \& Fabian 1991). As for the other broad line
parameters, our results are consistent with those reported above. In
particular, the disc emissivity is $q= 5.2^{+1.6}_{-1.3}$ and the
black hole spin is $a=0.64\pm 0.18$, confirming our previous estimate
and demonstrating that the reflection component does not affect the
relativistic blurring parameters significantly in this case.

\subsection{A self--consistent disc reflection model}

In order to attempt to model the data, and in particular the
reflection component, as self--consistently as possible, we use the
ionized reflection model from Ross \& Fabian (2005) where fluorescent
emission lines and reflection continuum are computed
together. Moreover, given the evidence for a broad Fe line, we
convolve the whole rest--frame reflection spectrum with the
relativistic kernel {\small{KYCONV}}, which extends the relativistic
blurring of the {\small{KYRLINE}} model to any intrinsic spectrum. The
reflection model we use is is computed by assuming a power law
continuum with fixed high--energy cut-off at 100~keV, but it is a pure
reflection model and does not include the illuminating continuum.  We
then model it by using a cut--off power law continuum with $E_c\equiv
100$~keV for consistency.

Ionized reflection naturally produces also a soft X--ray excess with
respect to the underlying power law continuum for a wide range of
ionization states. Hence, we start our analysis by removing the
blackbody component from our spectral model in the attempt of fitting
the whole broadband data with the most self--consistent model,
comprising just an absorbed power law (Galactic plus intrinsic ionized
absorption as before) plus disc reflection (keeping however the narrow
Fe K$\alpha$ line as before). The model has then 3 dof more than the
previous one. The ionization state of the reflector, which sets the
general spectral shape and the Fe line energy, is free to vary in the
fit, while we first fix the Fe abundance to the solar value.

The model provides a good representation of the data ($\chi^2=2728$
for 2758 dof) but the previous model was statistically slightly
better. The most noticeable difference is that the ionized absorber
tends to the lowest ionization state allowed by our table model
($\log\xi = -1$). We then replace it with a neutral intrinsic absorber
using the {\small{ZPHABS}} model and we obtain a significant
improvement ($\chi^2=2707$ for 2759 dof) measuring N$_{\rm
  H}^{\rm{z}}\simeq 4.5\times 10^{21}$~cm$^{-2}$. Given that the soft
excess is now described with disc reflection as opposed to the
phenomenological blackbody model discussed above, changes in the
absorber properties are not very surprising. However, the issue of
whether the intrinsic absorber is neutral or slightly ionized is not
completely settled in our opinion, and future high--resolution data
are required to resolve individual absorption features. Letting the Fe
abundance free top vary produces a further improvement with a final
result of ($\chi^2=2690$ for 2758 dof) with A$_{\rm{Fe}} \simeq 1.5$
times the solar value (see Table~2 for a summary of the
best--fitting parameters).
\begin{figure}
\begin{center}
\includegraphics[width=0.31\textwidth,height=0.45\textwidth,angle=-90]{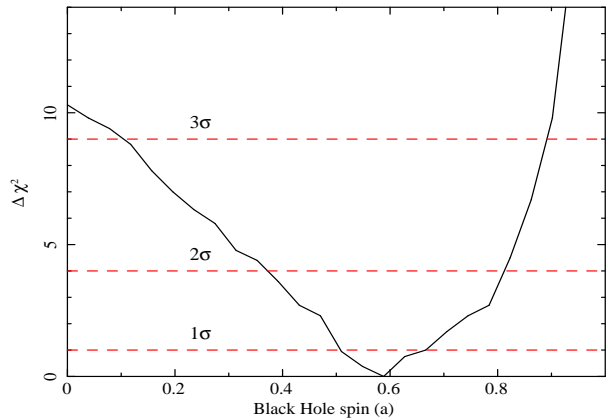}
\caption{Goodness--of--fit variation as a function of black hole
  spin. The horizontal lines mark $\Delta\chi^2=$1, 4, and 9,
  corresponding approximately to 1,2, and 3$\sigma$ confidence
  levels. A value of a=0 is rejected but only just at the 3$\sigma$
  level, while a maximally spinning black hole with a=0.998 can be
  rejected at the 5$\sigma$ level (not shown in the Figure for
  clarity, but corresponding to $\Delta\chi^2 =25.7$.)}
\label{goodness}
\end{center}
\end{figure}

\begin{table}
\begin{center}
  \caption{Best--fitting parameters for the power law plus
    relativistically blurred reflection model (also comprising an
    additional narrow Fe K$\alpha$ line).}
\begin{tabular}{lc}        
  \hline 
$\Gamma$ & $2.12 \pm 0.09 $ {\vspace{0.05cm}}\\ 
$N_H^{\rm{z},a}$ [cm$^{-2}$] & $4.5\pm 0.9${\vspace{0.05cm}}\\ 
$i$ [degrees]& $46\pm 4${\vspace{0.05cm}}\\ 
$q$ & $5.3^{+1.7}_{-1.4}$ {\vspace{0.05cm}}\\ 
$a$ & $0.6\pm 0.2$ {\vspace{0.05cm}}\\ 
$\xi_{\rm{ref}}$ [erg~cm~s$^{-1}$] & $40^{+70}_{-35}${\vspace{0.05cm}} \\ 
$A_{\rm{Fe}}$ [solar] & $1.5\pm 0.3${\vspace{0.05cm}}\\ 
$EW_{\rm{NL}}$ [eV]& $13\pm 9$ {\vspace{0.05cm}}\\ 
$\chi^2 /dof$ & 2690/2758\\ 
\hline
\end{tabular}
\end{center}
\label{tab2}
{\footnotesize{$^a$ Column density of the intrinsic absorber in units
    of $10^{21}$~cm$^{-2}$.}}
\end{table}

The ionization state of the reflector is such that the main emission
line in the Fe K regime is at 6.4~keV, as suggested already by
previous results (we measure $\xi = 40^{+70}_{-35}$~erg~cm~s$^{-1}$,
while the 6.7~keV Fe line starts to dominate for $\xi \geq
200$~erg~cm~s$^{-1}$). The relativistic blurring parameters are
consistent with previous results. In particular, we obtain an
emissivity index $q=5.3^{+1.7}_{-1.4}$ and a black hole spin $0.6 \pm
0.2$, while the inclination is $46^\circ \pm 5^\circ$. In
Fig.~\ref{goodness} we show the goodness--of--fit variation as a
function of black hole spin. We measure an intermediate spin of
$a\simeq 0.6$, but a Schwarzschild solution ($a=0$) can only be
rejected at just the 3$\sigma$ level. A maximally spinning Kerr black
hole is rejected with higher confidence. We must however point out
that the $a=0$ Schwarzschild solution requires a very steep emissivity
$q\geq 5.5$ which would associate all Fe emission with a narrow
annulus around the ISCO ($6~r_g$), with little contribution from larger
radii, which seems highly unlikely.

Despite the result being statistically slightly worse than the
previous one, we tend to consider our latest model superior from a physical
point of view. In fact, our latest model is particularly appealing
because it can describe the whole broadband X--ray spectrum with a
minimal set of well justified spectral components, namely: Galactic
and intrinsic absorption, a power law continuum, and its reflection
off the disc (producing the broad relativistic Fe line) and off
distant matter (as indicated by the narrow Fe line).

\subsection{X--ray variability}

\begin{figure}
\begin{center}
\includegraphics[width=0.33\textwidth,height=0.43\textwidth,angle=-90]{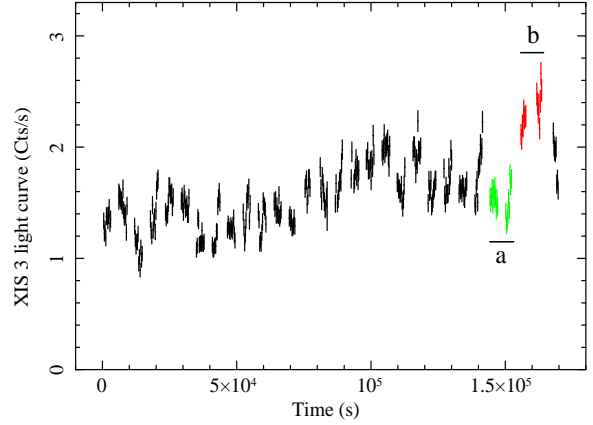}
\caption{The 0.5--10~keV XIS~3 background subtracted light curve of
  SWIFT~J2127.4+5654. Time bins are of 256~s and only fully exposed
  bins are used. Intervals a and b are labelled for reference (see
  text for details).}
\label{lc}
\end{center}
\end{figure}

So far, we have considered the time--averaged X--ray spectral
properties of SWIFT~J2127.4+5654. However, NLS1 galaxies are known for
their often extreme X--ray variability which is primarily attributed
 to a relatively small black hole mass. The black hole mass of
SWIFT~J2127.4+5654 has been estimated by Malizia et al. (2008) to be
of the order of $1.5\times 10^7~M_\odot$ (see Masetti et al 2006 for
the method). Although this is not a particularly low black hole mass,
short timescale variability should be present if, as indicated by our
spectral analysis, X--rays originate in the innermost regions of the
accretion flow. As a reference, the light--crossing time of $10~r_g$
for the given black hole mass corresponds to a timescale of $\sim
700$~s. In Fig.~\ref{lc} we show the XIS~3 light curve in the
0.5--10~keV band (bins of 256~s) which clearly exhibits variability of
more than 50 per cent in a few ks with a r.m.s. fractional variability
of $0.19\pm 0.08$.

We have performed time--resolved spectroscopy by selecting 30
intervals of 2--3~ks duration (i.e. the maximum length of
uninterrupted data, see Fig.~\ref{lc}) to search for spectral
variability and, in particular, to investigate any broad Fe line
variability with respect to the continuum. For simplicity, and to
avoid any soft excess complication, we restrict our analysis to the
3--10~keV band and we use only the XIS~3 because the lack of
sensitivity in the XIS~1 above 7--8~keV prevents us from defining the
hard X--ray continuum with the required accuracy. Given the relatively
poor quality of the time--resolved data, we use the simplest possible
spectral model comprising a power law continuum and two Gaussian
emission lines (one narrow and one broad) modified by an absorbing
column of neutral gas fixed at the time--averaged value. All
parameters are fixed to the best--fitting values from the
time--averaged spectrum, except the photon index, its normalization,
and the broad line intensity (see Table~1).Despite this restrictive
set of simplifying assumptions, the data are not of good enough
quality to claim any significant variation of the broad line
intensity. The continuum photon index is also consistent with being
the same throughout the observation. Considering longer time intervals
to improve the statistics would wash out most of the short timescale
flux variability, and we then refrain from performing such analysis on
the whole observation.

\begin{figure}
\begin{center}
\includegraphics[width=0.31\textwidth,height=0.45\textwidth,angle=-90]{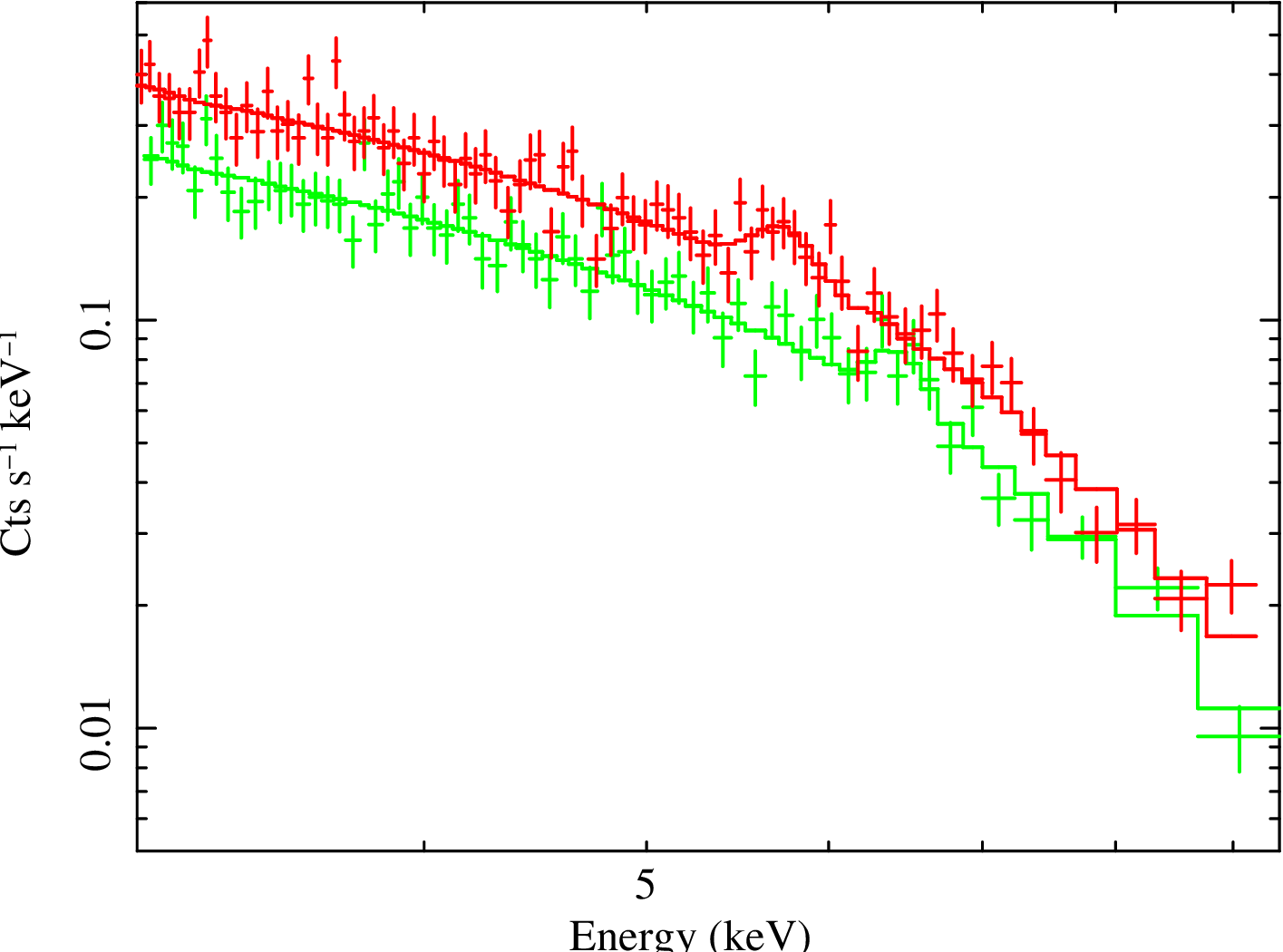}
{\vspace{0.27cm}}
\caption{The XIS~3 spectra for intervals a (bottom) and b (top) as
  defined in Fig.~\ref{lc}. The bulk of the broad Fe line seems to shift from
  $\sim$6.6~keV (a) to $\sim$5.8~keV (b) in about 10~ks.  }
\label{ab}
\end{center}
\end{figure}

The negative result could however be due, besides to the low quality
of the time--resolved data, to forcing the line parameters to be the
same, looking only for a line normalization change. To test for
changes in the line profile on short timescales and with a better
signal--to--noise, we have constructed spectra from intervals a and b
shown in Fig.~\ref{lc} (about 5--6~ks of net exposure each) and
applied the same model as before, but now letting all the broad Fe
line parameters free to vary. The resulting spectra and best--fitting
models are shown in Fig.~\ref{ab} and indicate the presence of a
spectral feature around 5.8~keV in the brighter spectrum only.  The
Gaussian lines in intervals a and b are resolved ($\sigma
\simeq$150~eV) have energies of $E^{\rm a} = 6.57 \pm 0.10$~keV and
$E^{\rm b} = 5.83 \pm 0.12$~keV respectively, and intensities of
$I^{\rm a} = 0.35\pm 0.15\times 10^{-4}$~ph~cm$^{-2}$~s$^{-1}$ and
$I^{\rm b} = 0.61\pm 0.20\times 10^{-4}$~ph~cm$^{-2}$~s$^{-1}$.

\begin{figure}
\begin{center}
\includegraphics[width=0.31\textwidth,height=0.45\textwidth,angle=-90]{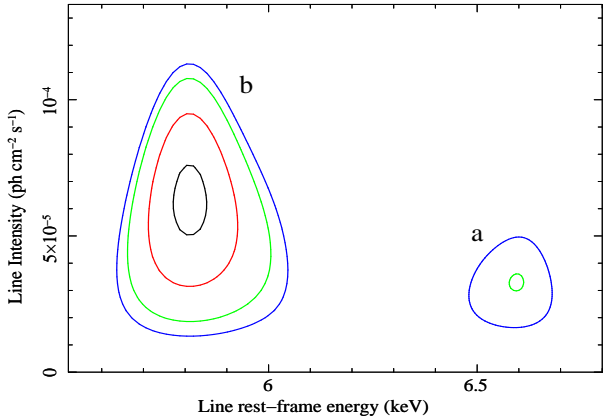}
\caption{Contours in the intensity--energy plane when a Gaussian
  filter is applied to the spectra of intervals a and b. The Gaussian
  has fixed width of $0.1$~keV, comparable with the XIS energy
  resolution, and is varied in the 4--8~keV energy and in the
  $0-4\times 10^{-4}$~ph~cm$^{-2}$~s$^{-1}$ intensity ranges. The
  contours represent, from outer-- to inner--most, $\Delta \chi^2 =
  -4.61$, $-6.17$, $-9.21$, and $-11.83$ with respect to a model when only
  a power law continuum and a narrow 6.4~keV Fe line are considered
  (see Miniutti \& Fabian 2006 for details).  }
\label{contab}
\end{center}
\end{figure}

The line parameters are then suggestive of some variability, at least
in the centroid energy which is significantly redder in the brighter
interval (b). This is visually shown in Fig.~\ref{contab} where we use
the method outlined in Miniutti \& Fabian (2006) and show the
statistical improvement obtained by running a Gaussian filter through
the data for intervals a and b when both spectra are modelled with the
power law plus narrow 6.4~keV Gaussian emission line only. While an
additional resolved line is confidently detected in interval b at
$\sim$5.8~keV, the detection of the line at $\sim$6.6~keV during
interval a is less significant (about 2.5$\sigma$). No line at
$\sim$6.6~keV is detected in interval b, as no line at $\sim$5.8~keV
is seen during interval a. However, if all broad line parameters are
forced to be the same in the two intervals, the statistical quality of
the fit becomes $\chi^2=145$ for 145 dof with respect to $\chi^2=135$
for 142 dof when the broad lines parameters are allowed to be
different. The line profile variability is then significant at the
$\sim$98 per cent confidence level only (we also point out that this
significance level is likely overestimated since we do not take into
account the number of trials for simplicity). We conclude that the
data quality is not high enough to study the reflection and Fe line
variability in detail, although some tentative evidence for line
profile variability is found on timescales $\leq$10~ks.

Attempting to explain the tentative variability we report is risky
(because of its low significance) and we do not want to
over--interpret our data. However, assuming that the variability is
real, the line appears to be more redshifted in the bright
interval. This is opposite to the behaviour predicted e.g. by the
light bending model (see Miniutti \& Fabian 2004) which, under the
assumption that flux variability is only due to general relativistic
effects, associates bright phases with narrower and less redshifted
lines. However, other interpretations may be viable and the 5.8~keV
line may be associated with the blue peak of an additional line
component rather than with the red wing of the broad line (which may
be too weak to be confidently detected in the time--resolved
intervals). For instance, the bright interval b could correspond to a
particularly strong X--ray flare above the disc, disconnected from the
overall variability trend, and illuminating only a limited off centre region of
the disc. The irradiated region then responds with Fe fluorescent line
emission with peak energy (and width) dictated by the location and
size of the illuminated spot (Dov{\v c}iak et al. 2004), as suggested
e.g. by Iwasawa et al. (1999) in the case of MCG--6-30-15 (see also
Iwasawa, Miniutti \& Fabian 2004 for an interesting case of flux and
energy modulation of such a feature in NGC~3516). The 5.8~keV feature
we observe during the bright interval b may than be due to a transient
flare above the receding part of the inner disc, which potentially
explains the relatively large observed redshift of the line.

\section{Discussion}

We detect with very high significance a broad Fe line in the
X--ray spectrum of the NLS1 galaxy SWIFT~J2127.4+5654. The line is
consistent with being relativistic and most likely originates in
reflection from the innermost regions of the accretion disc, close to
the central black hole. The good quality of our {\it Suzaku}
observation is sufficient to provide a measure of the black hole spin
in SWIFT~J2127.4+5654, a relatively rare opportunity in AGN studies.

The black hole in SWIFT~J2127.4+5654 appears to be a rotating Kerr
black hole with an intermediate spin of $0.6 \pm 0.2$. A non--rotating
Schwarzschild black hole ($a=0$) is statistically rejected but just at
the 3$\sigma$ level, while a maximally rotating Kerr one ($a=0.998$)
can be excluded at the 5$\sigma$ level (Fig.~\ref{goodness}). Since a
black hole having accreted a significant fraction of its mass should
spin at very high rate (Volonteri et al 2005) accretion in
SWIFT~J2127.4+5654 might proceed through short episodes with random
hole--disc angular momentum alignment. Accretion could then either add
or remove angular momentum depending on the relative orientation in
each accretion episode. The current accretion episode should be short
enough to prevent the hole from having accreted a significant fraction
of its mass thus reaching maximal or nearly maximal spin. If we assume
a typical efficiency of 0.1, and given that SWIFT~J2127.4+5654 has a
bolometric luminosity of $\sim 3.5\times 10^{44}$~erg~s$^{-1}$ (see
Table~1) for a black hole mass of $\sim 1.5\times 10^7~M_\odot$, the
black hole in SWIFT~J2127.4+5654 can double its mass in about
0.2-0.3~Gyr. This sets a conservative upper limit for the duration of
the accretion episode that is powering the AGN in SWIFT~J2127.4+5654.

We must point out that our spin measurement is based on the assumption
that the reflecting disc is truncated at the ISCO. This assumption is
reasonable, but can introduce a systematic error on the inferred spin
value if the disc is truncated near but not exactly at the ISCO,
i.e. if material in the plunging region beyond the ISCO can still act
as a reflector with low enough ionization to produce an observable
contribution to the reflection spectrum. Recent numerical simulations
suggest that the uncertainty in the reflector inner edge could at most
reduce the spin we have inferred by $\sim 0.2$ (Reynolds \& Fabian
2008). A lower black hole spin would only add strength to the previous
argument, favouring short--lived accretion episodes in
SWIFT~J2127.4+5654 as opposed to prolonged accretion. Given that
counter--rotating discs spin down black holes more efficiently than
co--rotating discs spin them up because gas particles leaving the ISCO
to plunge towards the black hole have larger angular momenta in the
counter--rotating case (see Bardeen 1970; Bardeen, Press \& Teukolsky
1972), the black hole spin distribution in this scenario should be
skewed towards low values. King, Pringle \& Hofmann (2008), who
investigated in detail the effect of chaotic accretion episodes on the
black hole spin, conclude indeed that the spin distribution for black
holes of $\sim 10^7~M_\odot$ should peak around $a=0.2-0.3$ with a
spread $\Delta a \simeq 0.2$, broadly consistent with our measurement,
especially if the correction due to the uncertainty in the reflector
inner edge is considered.

The second relevant result we have obtained concerns the shape of the
reflection emissivity profile. We measure a rather steep index
$q\simeq 5$ with a 90 per cent lower limit $q\geq 3.9$. A powerful and
centrally concentrated irradiating X--ray source is required to
achieve such steep profiles, possibly related to the magnetic
extraction of the black hole rotational energy via magnetic fields
(see e,g, Wilms et al. 2001 for a discussion). Relativistic light
bending may also contribute to focus the irradiating X--rays towards
the innermost disc regions, thus inducing an effective steeper
emissivity (Miniutti \& Fabian 2004). Here we stress that the
emissivity index $q$ and the black hole spin are somewhat degenerate
in spectral fitting: as shown in Fig.~\ref{spincont}, a steep
emissivity profile tends to force low values of spin (or large values
of the inner radius). Therefore, we detect a spin $a> 0$ despite (and
not because of) a steep emissivity. Moreover, from a physical point of
view, if the steep profile is indeed associated with the extraction of
the black hole rotational energy (plus light bending), steep
emissivity profiles should only be obtained for $a>0$. 

Rotational energy extraction tends to reduce the black hole spin, thus
representing a possible alternative to chaotic accretion as an
explanation for the intermediate spin we measure. If we assume
prolonged accretion (which tends to force a maximally spinning hole)
and rotational energy extraction (which tends to spin the hole down)
via the Blandford \& Znajek (BZ) mechanism (Blandford \& Znajek 1977),
the black hole spin evolution is different than with accretion alone,
and an equilibrium spin value is reached. It is interesting to note
that Moderski \& Sikora (1996), who analysed in some detail the
interplay between accretion and BZ mechanism in building up the black
hole spin, predict an equilibrium spin of $a_{\rm{eq}} ~\sim 0.6$ if
the current mass accretion rate (in terms of Eddington, see table~1)
of SWIFT~J2127.4+5654 and a standard $\alpha$--viscosity parameter of
0.1 are assumed. This estimate is actually remarkably consistent with
our spin measurement. We then consider it plausible that accretion in
SWIFT~J2127.4+5654 is prolonged, but that the spin is limited by
rotational energy extraction which may power the X--ray source. Such
occurence may explain simultaneously the spin value and the steep
emissivity profile we infer from our spectral analysis. The
intermediate spin of SWIFT~J2127.4+5654 is then consistent with two
competing scenarios, namely i) chaotic accretion episodes, or ii)
prolonged accretion plus rotational energy extraction via magnetic
mechanisms.

Tentative line profile variability is seen on timescales
$\leq$10~ks. Confirming such short--timescale variability in this and
other similar sources with broad lines would allow us to confidently
rule out any competing spectral model (such as complex absorption) and
to probe the relativistic regime around supermassive black
holes. Continuous data sets would be very helpful in assessing the
variability on short timescales (e.g. observations with {\it
  XMM--Newton}). On the other hand, a very significant improvement in
effective area at $\sim$6~keV with respect to current observatories is
necessary to probe the shortest timescales. Considering that the XIS~3
effective area is $\sim$220~cm$^2$ at 6~keV, $\sim$1~m$^2$ is required
to perform time--resolved spectroscopy of bright X--ray sources on
timescales shorter than the orbital one in the inner disc. Future
missions such as the International X--ray Observatory (IXO) will be
ideal to probe without ambiguity and with great accuracy the accretion
flow properties in the relativistic region.

\section*{Acknowledgments}

GM thanks the Spanish Ministerio de Ciencia e Innovaci\'on and CSIC
for support through a Ram\'on y Cajal contract. ACF thanks the Royal
Society for support. FP acknowledges support from the Italian Space
Agency (ASI) grant grant ASI/I/008/07.

\end{document}